\documentclass[preprint]{revtex4}%

\usepackage{amsfonts}
\usepackage{amsmath}
\usepackage{amssymb}
\usepackage{graphicx}%
\usepackage{xcolor}
\usepackage{subcaption}
\usepackage{graphicx}
\usepackage{float}

\DeclareMathOperator*{\argmax}{arg\,max}

\begin{document}
	\preprint{ }
	\title{Segmentation of quantum generated sequences by using the Jensen-Shannon divergence}
	
	\author{Marcelo Losada}
	\affiliation{Facultad de Matem\'atica, Astronom\'ia, F\'isica y Computaci\'on, Universidad Nacional de C\'ordoba, 5000 C\'ordoba, Argentina and Consejo Nacional de Investigaciones Cient\'ificas y T\'ecnicas, Argentina}
	
	\author{V\'ictor A. Penas}
	\affiliation{Facultad de Matem\'atica, Astronom\'ia, F\'isica y Computaci\'on, Universidad Nacional de C\'ordoba, 5000 C\'ordoba, Argentina and Consejo Nacional de Investigaciones Cient\'ificas y T\'ecnicas, Argentina}
	
\author{Federico Holik}
\affiliation{Instituto de F\'isica La Plata, CONICET, Diag.~113 e/63 y 64, 1900 La Plata, Argentina}
\vspace{10pt}

\author{Pedro W. Lamberti}
\affiliation{Facultad de Matem\'atica, Astronom\'ia, F\'isica y Computaci\'on, Universidad Nacional de C\'ordoba, 5000 C\'ordoba, Argentina and Consejo Nacional de Investigaciones Cient\'ificas y T\'ecnicas, Argentina}
\vspace{10pt}

\begin{abstract}
The Jensen-Shannon divergence has been successfully applied as a segmentation tool for symbolic sequences, that is to separate the sequence into subsequences with the same symbolic content.
In this work, we propose 
a method, based on the the Jensen-Shannon divergence, for segmentation of what we call \textit{quantum  generated sequences}, which consist in symbolic sequences generated from measuring a quantum system. 
For one-qubit and two-qubit systems, we show that the proposed method is adequate for segmentation. 
\end{abstract}

\startpage{1}
\maketitle

\section{Introduction}\label{sec: intro}

Quantum states are not directly observable. In some sense they share this property with probability distributions. In this last case we can only access to them through the frequency of occurrence of the possible values of the random variable which is described by the probability distribution. In this context a side problem is to distinguish two close probability distributions. The size of the statistical fluctuations will difficult the precise discrimination between them. Due to the statistical nature of quantum theory, there exists a corresponding problem of distinguishability between quantum states. 

To address these problems it is often necessary to define a distinguishability criterion, which requires to introduce distances on the probability distributions space and in the quantum states space, correspondingly. 

In 1982, C. Rao introduced an entropic distance between probability distributions \cite{Rao1982}, known as the Jensen Shannon divergence (JSD). In 1991, J. Lin studied a weighted version of the JSD that allows it to be applied in the context of the Bayesian inference theory \cite{Lin1991}. Recently an analogous version of the JSD has been defined between quantum states \cite{Lamberti2005, Lamberti2022}. Additionally, some generalizations of the JSD have been proposed by using alternative entropy functionals \cite{Tsallis1988, Lamberti2003, Azad2014}. 

The JSD divergence has several interesting properties among which the most significant is that its square root is a true metric on the set of probability distributions. The same property has been recently proven for the quantum version of the JSD \cite{Lamberti2008, Virosztek} 
and for their generalizations in the context of Tsallis and Rényi entropies \cite{Sra}.

The JSD has been exhaustively studied in the context
of the analysis of symbolic sequences;
in particular, a segmentation procedure that uses it has been applied  for symbolic sequences in \cite{Bernaola-Galvan1996, Bernaola-Galvan1999, Bernaola-Galvan2000, Azad2002, Grosse2002}. What we understand by segmentation is to separate the original sequence in subsequences with homogeneous compositional structure. 

In this work we propose a method based on the classical JSD, for the segmentation of what we call a \textit{quantum generated sequences}, which consists in symbolic sequences generated from measuring a quantum system. This proposal holds significant mathematical value as it represents a quantum generalization of the segmentation methods previously introduced in literature. For one-qubit and two-qubit systems, we show that the proposed method is adequate for segmentation.

The method presented here has been widely applied to the study of classical time series, both with chaotic and random behavior.
These applications could be adequately extended to the study of quantum system showing random and chaotic behavior. Moreover, our method could be usefully applied in the context of random quantum channels \cite{Karol, Karol2009}, taking into account the so-called Choi-Jamiolkowski isomorphism that sets up a correspondence  between quantum channels and  quantum states \cite{Choi}.

This paper is organized as follows. In Sec. \ref{sec: 2} we briefly review the main properties of the JSD. In Sec. \ref{sec: 3}, we introduce our new method, based on the classical JSD, for segmentation of quantum generated sequences.
For a matter of completeness, first, in Sec. \ref{sec: 3a}, we briefly describe the segmentation procedure based on the JSD for classical generated sequences. 
In Sec. \ref{sec: 4}, we consider simulations of quantum generated sequences coming from systems of one and two qubits. We applied the segmentation method proposed and we analyzed the obtained results.
Finally, Sec. \ref{sec: 5} concludes the article with a brief discussion of results obtained.

\section{Jensen-Shannon divergence}
\label{sec: 2}
Let us consider a discrete random variable $X$ with $n$ possible values $x_j$, with $1\leq j\leq n$. Let $\mathbf{P^1} = (p^1_1,\ldots, p^1_n)$ and $\mathbf{P^2} = (p^2_1,\ldots, p^2_n)$ be two probability distributions for $X$, with $p^1_j$ and $p^2_j$ the probability of occurrence of the event $X = x_j$ for the probability distributions $\mathbf{P^1}$ and $\mathbf{P^2}$, respectively. The Kullback–Leibler divergence (KLD) between  $\mathbf{P^1}$ and $\mathbf{P^2}$ is defined as 
\begin{equation}
D_{\text{KL}}(\mathbf{P^1} || \mathbf{P^2} ) = \sum_{j} p^1_j \ln \frac{p^1_j} {p^2_j}.
\end{equation}

The KLD is the natural dissimilarity measure aroused in information theory. However it has several drawbacks, mainly it is not always well defined and it is not a distance between probability distributions. To overcome these problems it was proposed a symmetrized version of the KLD, known as the JSD:

\begin{equation}
 \text{JSD}(\mathbf{P^1}|| \mathbf{P^2} ) =D_{\text{KL}}(\mathbf{P^1} || (\mathbf{P^2} +\mathbf{P^1})/2 ) + D_{\text{KL}}(\mathbf{P^2} || (\mathbf{P^2} +\mathbf{P^1})/2 ) .
 \end{equation}
 
This definition assumes that both probability distributions $\mathbf{P^1}$ and $\mathbf{P^2}$ are in the same footing, that is they have the same ``weights" ($\pi_1 = \pi_2 = 1/2$). If we consider arbitrary weights $\pi_1, \pi_2 \geq 0$ (with $\pi_1 + \pi_2 = 1$) for the probability distributions $\mathbf{P^1}$ and $\mathbf{P^2}$, a direct generalization  can be defined as follows 
\begin{equation}
 \text{JSD}^{(\pi_1, \pi_2 )}(\mathbf{P^1}, \mathbf{P^2}) = \pi_1 D_{\text{KL}}(\mathbf{P^1} ||\pi_1 \mathbf{P^1} + \pi_2 \mathbf{P^2} ) + \pi_2 D_{\text{KL}}(P^2 || \pi_1 \mathbf{P^1} + \pi_2 \mathbf{P^2} ).
 \end{equation}

The JSD can be expressed in terms of the Shannon entropy (H) as follows
\begin{equation}
 \text{JSD}^{(\pi_1, \pi_2 )}(\mathbf{P^1}, \mathbf{P^2} ) = H ( \pi_1 \mathbf{P^1} + \pi_2 \mathbf{P^2}) - \pi_1 H(\mathbf{P^1}) - \pi_2 H(\mathbf{P^2}).
 \end{equation}

The generalization of the JSD to $m$ probability distributions is straightforward,
\begin{equation}
 \text{JSD}^{(\pi_1, \ldots, \pi_m )}(\mathbf{P^1}, \ldots, \mathbf{P^m}) = H\Big(\sum_{i= 1}^m \pi_i \mathbf{P^i} \Big) - \sum_{i= 1}^m \pi_i H(\mathbf{P^i}).
 \end{equation}

The JSD possesses two key features that set it apart from the KLD. Firstly, it is always well-defined. Secondly, for the special case where $\pi_1 = \pi_2 = 1/2$, the JSD's square root can be considered as a metric.

\section{Sequence segmentation with Jensen-Shannon divergence}
\label{sec: 3}

In this section we introduce a method, based on the classical JSD, for segmentation of
what we call quantum generated sequences, which consists in symbolic sequences generated
from measuring a quantum system.
For a matter of completeness, first,  we briefly describe the segmentation procedure based on the JSD for classical generated sequences. The details can be found in \cite{Bernaola-Galvan1999, Grosse2002}. 

\subsection{Classical generated sequences}

\label{sec: 3a}

Let us consider a sequence of $n$ random variables $\left(X^i\right)_{i= 1}^{n}$, all with the same $m$ possible outcomes in the set of symbols $\mathcal{A} = \{a_1, \ldots, a_m  \}$. Each random variable have a probability distribution $\mathbf{P^{i}} =(p_1^{i}, \ldots, p_m^{i})$, with $p_j^{i}$ the probability of the event $X^i = a_j$. 

Any symbolic sequence  $\mathbf{s} = (s_1, \ldots , s_n)$, with each $s_i \in \mathcal{A}$, constitutes a possible realization of the random variables sequence 
$\left(X^i\right)_{i= 1}^{n}$. Since these symbolic sequences are generated from a sequence of classical random variables, we call them \textit{classical generated sequences}.

We are interested in classical generated sequences coming from sequences of independent random variables and with the following property: the first $l_1$ ($1 \leq l_1 < n$) random variables, $\left(X^i\right)_{i= 1}^{l_1}$, have the same probability  distribution $\mathbf{P^1}$ and the remain $l_2 = n- l_1$ random variables, $\left(X^i\right)_{i=l_1 +1}^{n}$, have the same probability distribution $\mathbf{P^2}$.
The index $i_{c} = l_1 +1$ will be called the changing index. 
These kind of sequences have only one changing index. However, the generalization to more changing indices is straightforward. 

The segmentation of a classical generated sequence with one changing index consists in finding the value $i_c$ of the symbolic sequence.
The Jensen-Shannon divergence was successfully applied as a tool for segmentation of this class of sequences \cite{Grosse2002} 

The general idea is the following. Given a classical generated sequence $\mathbf{s} = (s_1, \ldots, s_n )$, we consider a cursor that scrolls the sequence from $k= 2$ up to $k = n$. For each value of $k$, we split the sequence in two subsequences: $\mathbf{s^1}(k) = (s_1, \ldots, s_{k-1})$ and $\mathbf{s^2}(k) = (s_k, \ldots, s_n )$. 
For $\mathbf{s^1}(k)$ and $\mathbf{s^2}(k)$, we define the estimated  probability distributions $\mathbf{P^1}(k)$ and $\mathbf{P^2}(k)$, with their entries given by $p_j^1(k) = N_j^1(k)/(k-1)$ and $p_j^2(k) = N_j^2(k)/(n+1-k)$, with $N_j^i(k)$ the number of occurrences of the value $a_j$ in the sequence $\mathbf{s^i}(k)$.
Moreover, we define the associated weight $\pi_i(k)$ to each subsequence $\mathbf{s^i}(k)$ as follows,  $\pi_1(k) = (k-1)/n$ and $\pi_2(k) = (n+1- k)/n$.
Then, for each value of $k = 2, \ldots, n$, we compute $\text{JSD}^{(\pi_1(k),  \pi_2(k) )}(\mathbf{P^1}(k), \mathbf{P^2}(k))$.

Finally, we estimate the changing index $i_c$ as the position of the JSD, i.e., 
\begin{equation}
   \hat{i}_c = \argmax_{1<  k \leq n} ~ \text{JSD}^{(\pi_1(k),  \pi_2(k) )}(\mathbf{P^1}(k), \mathbf{P^2}(k)).
\end{equation}

This segmentation procedure was shown to be successful in several context (see \cite{Azad2014} and references therein). 

In the next subsection, we  introduce the notion of quantum generated sequences and we show how the segmentation method based on the JSD can be generalized to these sequences.

\subsection{Quantum generated sequences}
\label{quantum generated sequences}

In the quantum case, instead of having a sample space, random variables and probability distributions, we have a Hilbert space, quantum observables, represented by Hermitian operators, and quantum states, represented by density matrices. We have to replace the sequences of random variables with sequences of quantum observables. 

Let us consider a Hilbert space $\mathcal{H}$ with dimension $m$, and a set of Hermitian operators acting on $\mathcal{H}$ with only non-degenerate eigenvalues, $\mathcal{O} = \{ O^1, \ldots, O^d\}$.
Each Hermitian operator $O^{r}$ ($1 \leq r \leq d$) 
has a spectral decomposition, $O^{r} = \sum_j a^r_j\Pi^r_j$,
where $\mathcal{A}^r = \{ a_1^r, \ldots, a_m^r\}$ is set of  possible outcomes of the observable $O^{r}$
and $\{\Pi^r_j\}_{j= 1}^{m}$ are orthogonal projectors satisfying $\sum_j \Pi^r_j = I$ (with $I$ the identity operator of $\mathcal{H}$).

Now, let us consider a sequence $\mathbf{O}$ of $n$ observables chosen from the set $\mathcal{O}$, i.e., 
$\mathbf{O} = \left( O^{r_i} \right)_{i= 1}^{n}$, with $r_i \in\{ 1, \ldots, d\}$, and a sequence of $n$ density matrices $\left( \rho_{i} \right)_{i= 1}^{n}$.
If, for each state $\rho_i$, we measure the observable $O^{r_i}$, we can obtain any outcome $a_j^{r_i} \in \mathcal{A}^{r_i}$ with probability $\text{tr}(\rho_i \Pi_j^{r_i})$.

Any sequence  $\mathbf{s} = (s_1, \ldots , s_n)$, with each $s_i \in \mathcal{A}^{r_i}$, constitutes a possible realization of the sequence of observables $\mathbf{O}=\left( O^{r_i} \right)_{i= 1}^{n}$. Since these sequences are generated from a sequence of quantum observables, we call them \textit{quantum generated sequences}.
We are interested in quantum generated sequences 
coming from 
sequences of density matrices such that 
the first $l_1$ ($1 \leq l_1 < n$) are the same (i.e., $\rho_i = \rho_1$ for $1 \leq i \leq l_1$) and the remain $l_2 = n- l_1$ are also the same ($\rho_i = \rho_2$ for $l_1 < i \leq n$).
Again, $i_{c} = l_1 +1$ will be called the changing index. 
These kind of sequences have only one changing index. However, the generalization to more changing indices is straightforward. 

In what follows we propose a method based on the Jensen-Shannon divergence for the segmentation of a quantum generated sequence with one changing index.
Given a quantum generated sequence $\mathbf{s} = (s_1, \ldots, s_n )$, we consider a cursor that scrolls the sequence from $k= 2$ up to $k = n$. For each value of $k$, we split the sequence in two subsequences: $\mathbf{s^1}(k) = (s_1, \ldots, s_{k-1})$ and $\mathbf{s^2}(k) = (s_k, \ldots, s_n )$. 
For $\mathbf{s^1}(k)$ and $\mathbf{s^2}(k)$, and for each observable $O^r \in \mathcal{O}$, we define the estimated probability distributions $\mathbf{P^{1,r}}(k)$ and $\mathbf{P^{2,r}}(k)$ as follows:
\begin{equation}
    p_j^{1,r}(k) = N_j^{1,r}(k)/N^{1,r}(k),  ~~~~~ \text{and} ~~~~~ p_j^{2,r}(k) = N_j^{2,r}(k)/N^{2,r}(k), 
\end{equation}
with $N_j^{i,r}(k)$ the number of occurrences of the value $a_j^r$ in the sequence $\mathbf{s^i}(k)$ and $N^{i,r}(k)$ the number of occurrences of an outcome coming from the observable $O^r$ in the sequence $\mathbf{s^i}(k)$. In case $N^{i,r}(k) = 0$, there are no measurements of the observable $O^r$, therefore, it is not possible to define the probability distribution $\mathbf{P^{i,r}}(k)$.

Then, for each value of $k = 2, \ldots, n$ and $r= 1, \ldots, d$, we compute $\text{JSD}^{r}(k) = \text{JSD}^{(\pi_1(k),  \pi_2(k) )}(\mathbf{P^{1,r}}(k), \mathbf{P^{2,r}}(k))$. In case $\mathbf{P^{1,r}}(k)$ or $\mathbf{P^{1,r}}(k)$ are not defined, we stipulate that $\text{JSD}^{r}(k) = 0$.
Then, we obtain the maximum over all $r \in \{1, \ldots, d\}$,
\begin{equation}
   \text{JSD}^{max} (k) = \max_{1\leq r \leq d} \text{JSD}^{r} (k).
\end{equation}

Finally, we estimate the changing index $i_c$ as the position of the maximum of $\text{JSD}^{max}(k)$, i.e., \begin{equation}
   \hat{i}_c = \argmax_{1<  k \leq n} ~  \text{JSD}^{max} (k).
\end{equation}

In the next section we applied this method to the segmentation of quantum generated sequences from qubit systems.

\section{Examples}

\label{sec: 4}

In this section we consider simulations of quantum generated sequences coming from systems of one and two qubits. We applied the segmentation method proposed in Sec. \ref{quantum generated sequences} and we analyzed the obtained results.

\subsection{One-qubit system}

\label{sec: 4a}

We considered a one-qubit system and the set of Hermitian operators $\mathcal{O} = \{ X, Y, Z\}$, with $X$, $Y$ and $Z$ the Pauli matrices.
Moreover, we considered a sequence of $n = 2000$ observables, $\mathbf{O}_{X,Y,Z}$, given by successively alternating the Pauli observables, i.e.,   
$\mathbf{O}_{X,Y,Z} = \left(X, Y, Z, X, Y, Z, \ldots, X \right)$. 
For the first $l_1 = 1000$ observables, the system was in the state $|\psi_1 \rangle = \frac{1}{\sqrt{3}}|0\rangle + \frac{2i}{\sqrt{3}}|1\rangle$, and for the remain $l_2 = 1000$ observables, the system was in the state $|\psi_2 \rangle = \frac{1}{\sqrt{9}}|0\rangle + \frac{8i}{\sqrt{9}}|1\rangle$.

We simulated the measurements of the sequence of observables $\mathbf{O}_{X,Y,Z}$. The first $l_1 = 1000$ measurements were simulated using the state $|\psi_1 \rangle$ and the remain $l_2 = 1000$ measurements were simulated using the state $|\psi_2 \rangle$. The results of the simulated measurements constitutes a quantum generated sequence $\mathbf{s}$.

We applied the method presented in the Subsection \ref{quantum generated sequences} for the
segmentation of the sequence $\mathbf{s}$. 
We considered a cursor that scrolls the sequence from $k= 2$ up to $k = 2000$. For each value of $k$, we splited the sequence in two subsequences: $\mathbf{s^1}(k) = (s_1, \ldots, s_{k-1})$ and $\mathbf{s^2}(k) = (s_k, \ldots, s_n )$. 
For each $\mathbf{s^i}(k)$ ($i = 1, 2$) and for each Pauli matrix, we computed the following  estimated probability distributions:
\begin{align}\mathbf{P^{i,X}}(k)&= \Big(\frac{N_{+}^{i,X}(k)}{N^{i,X}(k)}, \frac{N_{-}^{i,X}(k)}{N^{i,X}(k)} \Big), \\
\mathbf{P^{i,Y}}(k)&=  \Big(\frac{N_{+}^{i,Y}(k)}{N^{i,Y}(k)}, \frac{N_{-}^{i,Y}(k)}{N^{i,Y}(k)} \Big), \\
\mathbf{P^{i,Z}}(k)&= \Big(\frac{N_{+}^{i,Z}(k)}{N^{i,Z}(k)}, \frac{N_{-}^{i,Z}(k)}{N^{i,Z}(k)} \Big).
\end{align}

The associated weight $\pi_i(k)$ to each subsequence $\mathbf{s^i}(k)$ was given by $\pi_1(k) = (k-1)/n$ and $\pi_2(k) = (n+1- k)/n$.
Then, for each value of $k = 2, \ldots, n$ and $r= X, Y, Z$, we computed $\text{JSD}^{r}(k) = \text{JSD}^{(\pi_1(k),  \pi_2(k) )}(\mathbf{P^{1,r}}(k), \mathbf{P^{2,r}}(k))$. 
Then, we obtained the maximum over all $r \in \{X, Y, Z\}$: 
$\text{JSD}^{max} (k) = \max_{r \in \{X, Y, Z\}} \text{JSD}^{r} (k)$. Finally, 
we estimated the changing index $i_c$ as the position of the maximum of $\text{JSD}^{max}(k)$, i.e., $\hat{i}_c = \argmax_{1<  k \leq n} ~  \text{JSD}^{max} (k)$.

\begin{figure}[H]
\centering
\includegraphics[width=0.4\textwidth]{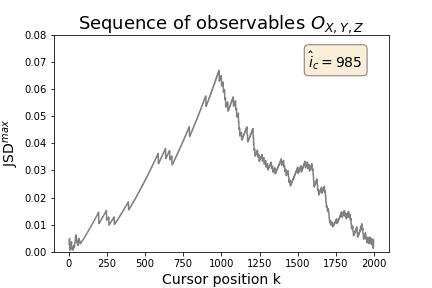}
\caption{\small{For the sequence of observables $\mathbf{O}_{X,Y,Z}$, and measurements simulated with states $|\psi_1 \rangle$ and $|\psi_2 \rangle$, we plot the value of $\text{JSD}^{max} (k)$ for each position of the cursor $k$. The position of the maximum is $\hat{i}_c =  985$, a good estimation of the changing index of the generated sequence.}}
\label{f:fig1}
\end{figure}

In Fig. \ref{f:fig1}, for each position of the cursor, we plotted the value of $\text{JSD}^{max} (k)$. The position of the maximum $\hat{i}_c$ is $k = 985$, a good estimation of the changing index $i_c = 1001$ of the generated sequence $\mathbf{s}$.

We repeated the same process, but considering different sequences of observables.
We considered sequences with only one Pauli observable, i.e,  $\mathbf{O}_{X} = \left(X,  \ldots, X \right)$, $\mathbf{O}_{Y} = \left(Y,  \ldots, Y \right)$ and $\mathbf{O}_{Z} = \left(Z,  \ldots, Z \right)$.
Also, we considered sequences 
given by successively alternating two Pauli observables, for example $\mathbf{O}_{X,Y} = \left(X, Y,   \ldots, X, Y \right)$, $\mathbf{O}_{X,Z} = \left(X, Z,   \ldots, X, Z \right)$ and  $\mathbf{O}_{Y, Z} = \left(Y, Z,   \ldots, Y, Z \right)$.
We simulated 
the measurements for each sequence of observables. As in the first example, the first $l_1 = 1000$ measurements were simulated using the state $|\psi_1 \rangle$ and the remain $l_2 = 1000$ measurements were simulated using the state $|\psi_2 \rangle$.

\begin{figure}[htb]
 \centering
   \begin{subfigure}[b]{.30\textwidth}
 \centering
  \includegraphics[width=1.0\linewidth]{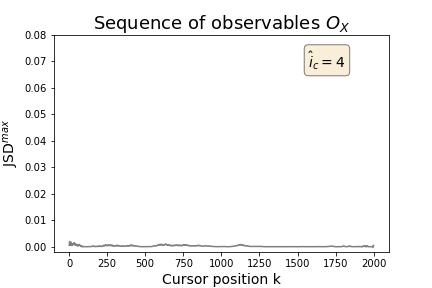}   
 \end{subfigure}\hfil
 \begin{subfigure}[b]{.30\textwidth}
 \centering
  \includegraphics[width=1.0\linewidth]{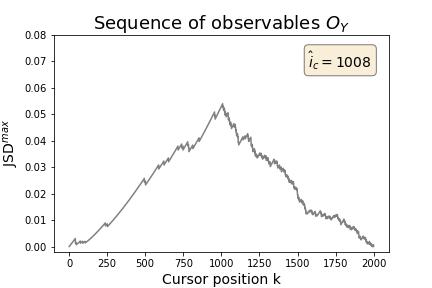}   
 \end{subfigure}\hfil
\begin{subfigure}[b]{.30\textwidth}  \label{2 crecientes}
  \centering
    \includegraphics[width=1.0\linewidth]{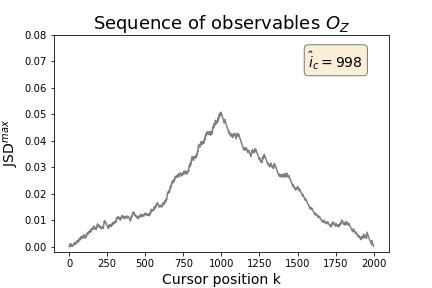}
 
\end{subfigure}\hfil
\begin{subfigure}[b]{.3\textwidth}
 \centering
  \includegraphics[width=1.0\linewidth]{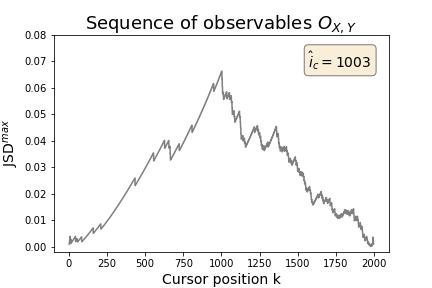}   
 \end{subfigure}\hfil
\begin{subfigure}[b]{.3\textwidth}
  \centering
  \includegraphics[width=1.0\linewidth]{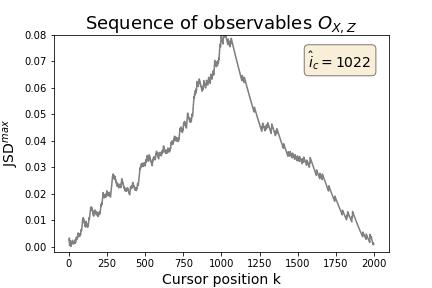}    
\end{subfigure}\hfil
\begin{subfigure}[b]{.3\textwidth}
  \centering
    \includegraphics[width=1.0\linewidth]{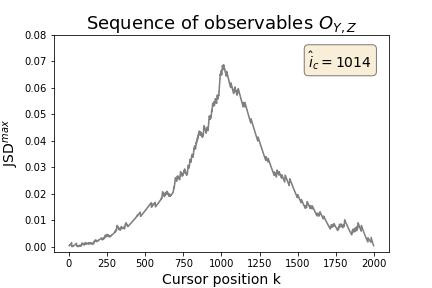}
 \end{subfigure}\hfil

\caption{For sequences of observables $\mathbf{O}_{X}$, $\mathbf{O}_{Y}$, $\mathbf{O}_{Z}$, $\mathbf{O}_{X,Y}$, $\mathbf{O}_{X, Z}$,  $\mathbf{O}_{Y,Z}$, and measurements simulated with states $|\psi_1 \rangle$ and $|\psi_2 \rangle$, we plot the value of $\text{JSD}^{max} (k)$ for each position of the cursor $k$. We observe that for the all sequences of observables, except  $\mathbf{O}_{X}$, the position of the maximum $\hat{i}_c$ is a good estimator of the changing index of the generated sequences.}
\label{f:fig2}
\end{figure}

In Fig. \ref{f:fig2} we present the obtained results. 
We can see that for the all sequences of observables, except  $\mathbf{O}_{X}$, the position of the maximum $\hat{i}_c$ is a good estimator of the changing index $i_c= 1001$ of the generated sequence.
For the sequence $\mathbf{O}_{X}$ this is not the case. The reason is that the two states $|\psi_1 \rangle$ and $|\psi_2 \rangle$ have the same probability distribution associated with the observable $X$, therefore, it is necessary to consider other basis to distinguish both states. 

\begin{figure}[H]
\centering
\includegraphics[width=0.4\textwidth]{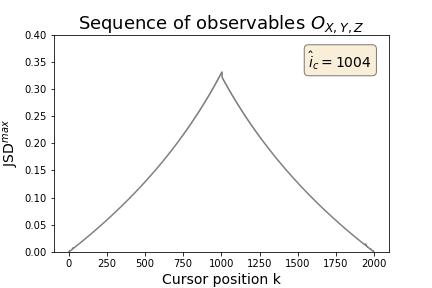}
\caption{\small{For the sequence of observables $\mathbf{O}_{X,Y,Z}$, and measurements simulated with states $\rho_1$ and $\rho_2$, we plot the value of $\text{JSD}^{max} (k)$ for each position of the cursor $k$. The position of the maximum $\hat{i}_c$ is $k = 1004$, a good estimation of the changing index of the generated sequence.}}
\label{f:fig3}
\end{figure}

\begin{figure}[htb]
 \centering
   \begin{subfigure}[b]{.30\textwidth}
 \centering
  \includegraphics[width=1.0\linewidth]{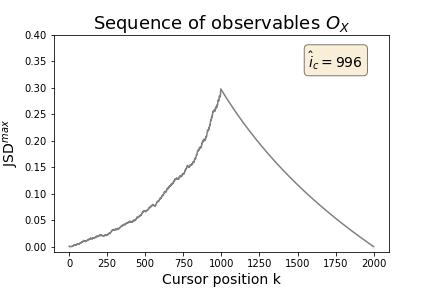}   
 \end{subfigure}\hfil
 \begin{subfigure}[b]{.30\textwidth}
 \centering
  \includegraphics[width=1.0\linewidth]{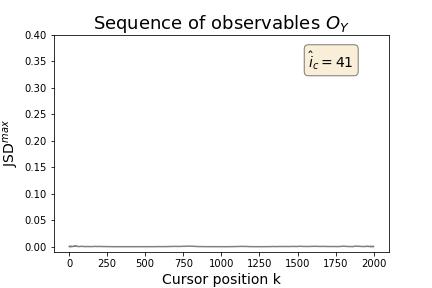}   
 \end{subfigure}\hfil
\begin{subfigure}[b]{.30\textwidth}  \label{2 crecientes}
  \centering
    \includegraphics[width=1.0\linewidth]{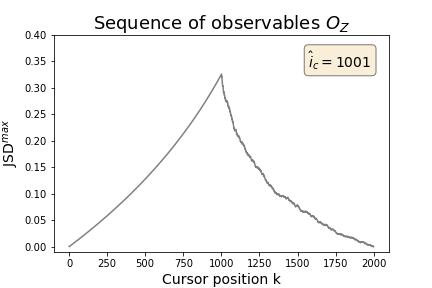}
 
\end{subfigure}\hfil
\begin{subfigure}[b]{.3\textwidth}
 \centering
  \includegraphics[width=1.0\linewidth]{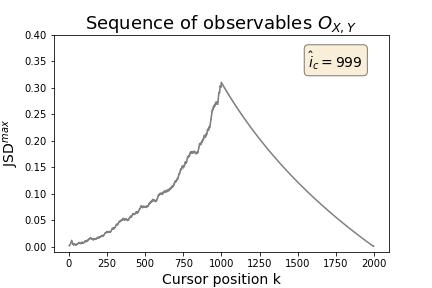}   
 \end{subfigure}\hfil
\begin{subfigure}[b]{.3\textwidth}
  \centering
  \includegraphics[width=1.0\linewidth]{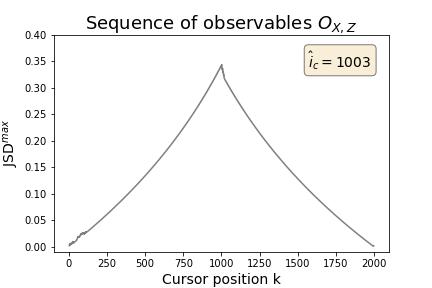}    
\end{subfigure}\hfil
\begin{subfigure}[b]{.3\textwidth}
  \centering
    \includegraphics[width=1.0\linewidth]{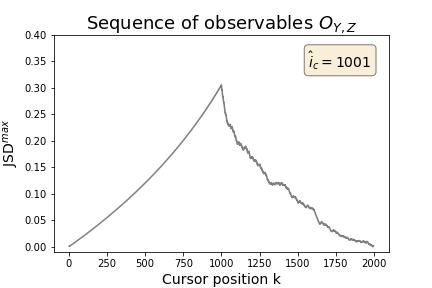}
 \end{subfigure}\hfil
\caption{For sequences of observables $\mathbf{O}_{X}$, $\mathbf{O}_{Y}$, $\mathbf{O}_{Z}$, $\mathbf{O}_{X,Y}$, $\mathbf{O}_{X, Z}$,  $\mathbf{O}_{Y,Z}$, and measurements simulated with states $\rho_1$ and $\rho_2$, we plot the value of $\text{JSD}^{max} (k)$ for each position of the cursor $k$. We observe that for the all sequences of observables, except  $\mathbf{O}_{Y}$, the position of the maximum $\hat{i}_c$ is a good estimator of the changing index of the generated sequences.}
\label{f:fig4}
\end{figure}

We repeated the study for the states 
$\rho_1 = |0 \rangle \langle 0|$ and $\rho_2 =I/2$. In Fig. \ref{f:fig3}, we plot the results obtained for the sequence of observables $\mathbf{O}_{X, Y, Z}$. Again we obtained that the position of the maximum $\hat{i}_c$ is a good estimation of the changing index of the generated sequence.
In Fig. \ref{f:fig4}, we plot the results obtained for sequences of observables $\mathbf{O}_{X}$, $\mathbf{O}_{Y}$, $\mathbf{O}_{Z}$ $\mathbf{O}_{X, Y}$, $\mathbf{O}_{X, Z}$, $\mathbf{O}_{Y, Z}$.
In this case, we can see that for the all sequences of observables, except  $\mathbf{O}_{Y}$, the position of the maximum $\hat{i}_c$ is a good estimator of the changing index of the generated sequence.
For the sequence $\mathbf{O}_{Y}$ this is not the case. The reason is that the two states $\rho_1$ and $\rho_2$ have the same probability distribution associated with the observable $Y$, therefore, it is necessary to consider other basis to distinguish both states. 

\subsection{Two-qubit system}

\label{sec: 4b}

In this subsection we present the application of the segmentation method  proposed in Sec. \ref{quantum generated sequences} to a system of two qubits.
In this case we consider 
the set of Hermitian operators $\mathcal{O} = \{ A \otimes B\}_{A,B = X, Y, Z}$, with $X$, $Y$ and $Z$ the Pauli matrices.

In this case we considered a sequence of $n = 2000$ observables, $\mathbf{O}_{XX,YY,ZZ}$, given by $\left(X \otimes X, Y \otimes Y, Z \otimes Z, X \otimes X, Y \otimes Y, Z \otimes Z, \ldots, X \otimes X \right)$. 
For the first $l_1 = 1000$ observables, the system was in the state $|\psi_1 \rangle = |00\rangle $, and for the remain $l_2 = 1000$ observables, the system was in the state $|\psi_2 \rangle = \frac{1}{\sqrt{2}}|00\rangle + \frac{1}{\sqrt{2}}|11\rangle$.
The results of the simulated measurements constitutes a quantum generated sequence $\mathbf{s}$.

We considered a cursor that scrolls the sequence from $k= 2$ up to $k = 2000$. For each value of $k$, we splited the sequence in two subsequences: $\mathbf{s^1}(k) = (s_1, \ldots, s_{k-1})$ and $\mathbf{s^2}(k) = (s_k, \ldots, s_n )$. 
For each $\mathbf{s^i}(k)$ ($i = 1, 2$) and for each operator $X\otimes X$, $Y\otimes Y$ and $Z\otimes Z$,   we computed the following  estimated probability distributions:
\begin{align}\mathbf{P^{i,XX}}(k)&= \Big(\frac{N_{+}^{i,XX}(k)}{N^{i,XX}(k)}, \frac{N_{-}^{i,XX}(k)}{N^{i,XX}(k)} \Big), \\
\mathbf{P^{i,YY}}(k)&=  \Big(\frac{N_{+}^{i,YY}(k)}{N^{i,YY}(k)}, \frac{N_{-}^{i,YY}(k)}{N^{i,YY}(k)} \Big), \\
\mathbf{P^{i,ZZ}}(k)&= \Big(\frac{N_{+}^{i,ZZ}(k)}{N^{i,ZZ}(k)}, \frac{N_{-}^{i,ZZ}(k)}{N^{i,ZZ}(k)} \Big).
\end{align}

$\mathbf{P^{i,X}}(k)$, $\mathbf{P^{i,Y}}(k)$, $\mathbf{P^{i,Z}}(k)$.

The associated weight $\pi_i(k)$ to each subsequence $\mathbf{s^i}(k)$ was given by $\pi_1(k) = (k-1)/n$ and $\pi_2(k) = (n+1- k)/n$.
Then, for each value of $k = 2, \ldots, n$ and $r= XX, YY, ZZ$, we computed $\text{JSD}^{r}(k) = \text{JSD}^{(\pi_1(k),  \pi_2(k) )}(\mathbf{P^{1,r}}(k), \mathbf{P^{2,r}}(k))$. 
Then, we obtained the maximum over all $r \in \{X, Y, Z\}$: 
$\text{JSD}^{max} (k) = \max_{r \in \{XX, YY, ZZ\}} \text{JSD}^{r} (k)$. Finally, 
we estimated the changing index $i_c$ as the position of the maximum of $\text{JSD}^{max}(k)$, i.e., $\hat{i}_c = \argmax_{1<  k \leq n} ~  \text{JSD}^{max} (k)$.

\begin{figure}[H]
\centering
\includegraphics[width=0.4\textwidth]{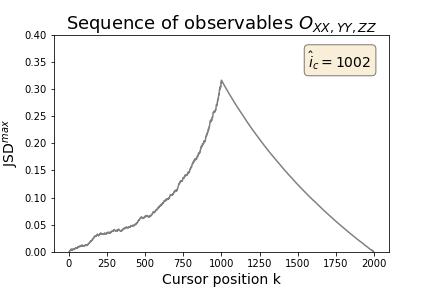}
\caption{\small{For the sequence of observables $\mathbf{O}_{XX,YY,ZZ}$, and measurements simulated with states $|\psi_1 \rangle$ and $|\psi_2 \rangle$, we plot the value of $\text{JSD}^{max} (k)$ for each position of the cursor $k$. The position of the maximum $\hat{i}_c$ is $k = 1000$, a good estimation of the changing index of the quantum generated sequence.}}
\label{f:fig5}
\end{figure}

In Fig. \ref{f:fig5}, for each position of the cursor, we plotted the value of $\text{JSD}^{max} (k)$. The position of the maximum $\hat{i}_c$ is $k = 1000$, a good estimation of the changing index $i_c = 1001$ of the quantum generated sequence.

We repeated the same process, but considering different sequences of observables.
We considered the following  sequences $\mathbf{O}_{XX} = \left(X\otimes X,  \ldots, X\otimes X \right)$, $\mathbf{O}_{XY} = \left(X\otimes Y,  \ldots, X\otimes Y \right)$, $\mathbf{O}_{XZ} = \left(X\otimes Z,  \ldots, X\otimes Z \right)$, $\mathbf{O}_{YY} = \left(Y\otimes Y,  \ldots, Y\otimes Y \right)$, $\mathbf{O}_{YZ} = \left(Y\otimes Z,  \ldots, Y\otimes Z \right)$, and $\mathbf{O}_{ZZ} = \left(Z\otimes Z,  \ldots, Z\otimes Z \right)$.
We simulated 
the measurements for each sequence of observables. As in the first example, the first $l_1 = 1000$ measurements were simulated using the state $|\psi_1 \rangle$ and the remain $l_2 = 1000$ measurements were simulated using the state $|\psi_2 \rangle$.

\begin{figure}[htb]
 \centering
   \begin{subfigure}[b]{.30\textwidth}
 \centering
  \includegraphics[width=1.0\linewidth]{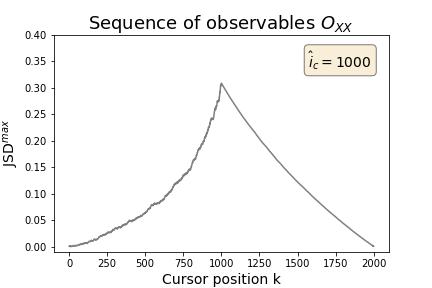}   
 \end{subfigure}\hfil
 \begin{subfigure}[b]{.30\textwidth}
 \centering
  \includegraphics[width=1.0\linewidth]{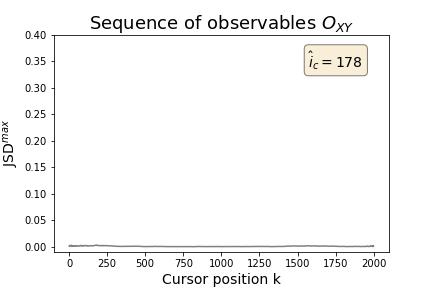}   
 \end{subfigure}\hfil
\begin{subfigure}[b]{.30\textwidth}  \label{2 crecientes}
  \centering
    \includegraphics[width=1.0\linewidth]{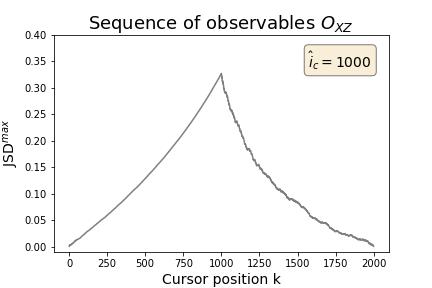}
 
\end{subfigure}\hfil
\begin{subfigure}[b]{.3\textwidth}
 \centering
  \includegraphics[width=1.0\linewidth]{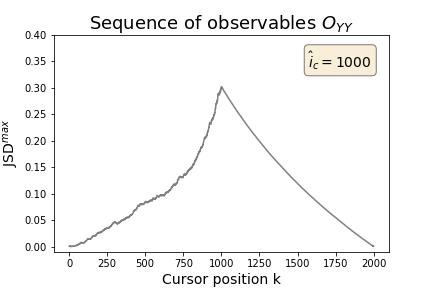}   
 \end{subfigure}\hfil
\begin{subfigure}[b]{.3\textwidth}
  \centering
  \includegraphics[width=1.0\linewidth]{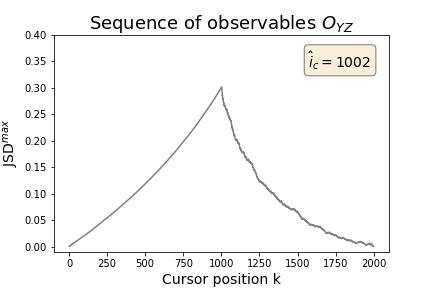}    
\end{subfigure}\hfil
\begin{subfigure}[b]{.3\textwidth}
  \centering
    \includegraphics[width=1.0\linewidth]{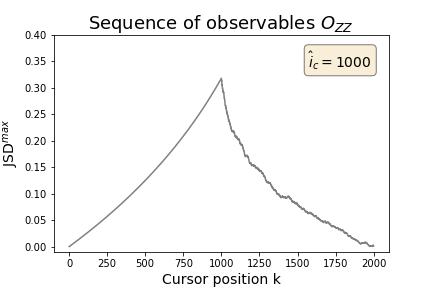}
 \end{subfigure}\hfil

\caption{For sequences of observables $\mathbf{O}_{XX}$, $\mathbf{O}_{XY}$, $\mathbf{O}_{XZ}$, $\mathbf{O}_{YY}$, $\mathbf{O}_{YZ}$,  $\mathbf{O}_{ZZ}$, and measurements simulated with states $|\psi_1 \rangle$ and $|\psi_2 \rangle$, we plot the value of $\text{JSD}^{max} (k)$ for each position of the cursor $k$. We observe that for the all sequences of observables, except  $\mathbf{O}_{XY}$, the position of the maximum $\hat{i}_c$ is a good estimator of the changing index of the generated sequences.}
\label{f:fig6}
\end{figure}

In Fig. \ref{f:fig6} we present the obtained results. 
We can see that for the all sequences of observables, except  $\mathbf{O}_{XY}$, the position of the maximum $\hat{i}_c$ is a good estimator of the changing index $i_c= 1001$ of the generated sequence.
For the sequence $\mathbf{O}_{XY}$ this is not the case. The reason is that the two states $|\psi_1 \rangle$ and $|\psi_2 \rangle$ have the same probability distribution associated with the observable $X \otimes Y$, therefore, it is necessary to consider other basis to distinguish both states.

\section{Conclusions}
\label{sec: 5}

The Jensen-Shannon divergence has been successfully applied as a segmentation tool for symbolic
sequences, that is to separate the sequence into subsequences with the same symbolic content.

In this work we introduced the notion of quantum generated sequences, which consists in symbolic sequences generated from measuring a quantum system. This definition is a generalization based on the classical generated sequences, which are symbolic sequences generated from measuring a classical system.

For quantum generated sequences, we proposed a segmentation method, based on the classical JSD.
This proposal holds significant mathematical value as
it represents a quantum generalization of the segmentation methods previously proposed
in literature. 
We showed for one-qubit and two-qubit systems that the proposed method is adequate for segmentation.

Moreover, our method could be useful in the context of random quantum channels \cite{Karol}, taking into account the Choi-Jamiolkowski isomorphism
 among quantum channels and mixed quantum states \cite{Choi}.

\section*{Acknowledgments}
M.L., V.P., F.H. and P.W.L. acknowledges partial support from CONICET, Argentina. F.H. was partially funded by PICT- 2019-01272: Estructuras l\'ogicas y algebraicas vinculadas al procesamiento
de la informaci\'on cu\'antica; and by the project “Per un’estensione semantica della Logica Computazionale Quantistica- Impatto teorico e ricadute implementative”, Regione Autonoma della Sardegna, (RAS: RASSR40341), L.R. 7/2017, annualità 2017- Fondo di Sviluppo e Coesione (FSC) 2014–2020.

\end{document}